\documentclass[twocolumn]{aastex631} 

\usepackage{amsmath}
\usepackage{rotating}
\usepackage{float}

\begin{document}

\title{Delivery of DART Impact Ejecta to Mars and Earth: Opportunity for Meteor Observations}

\author[0000-0002-7257-2150]{Eloy Peña-Asensio}
\affiliation{Department of Aerospace Science and Technology, Politecnico di Milano, Via La Masa 34, Milano, 20156, Lombardia, Italy}
\affiliation{Departament de Química, Universitat Autònoma de Barcelona 08193 Bellaterra, Catalonia, Spain}
\affiliation{Institut de Ciències de l’Espai (ICE, CSIC) Campus UAB, C/ de Can Magrans s/n, 08193 Cerdanyola del Vallès, Catalonia, Spain}

\author[0000-0002-5666-8582]{Michael Küppers}
\affiliation{European Space Agency, European Space Astronomy Centre, Camino bajo del Castillo S/N Urbanización Villafranca del Castillo, 28692 Villanueva de la Cañada, Madrid, Spain}

\author[0000-0001-8417-702X]{Josep M. Trigo-Rodríguez}
\affiliation{Institut de Ciències de l’Espai (ICE, CSIC) Campus UAB, C/ de Can Magrans s/n, 
08193 Cerdanyola del Vallès, Catalonia, Spain}
\affiliation{Institut d’Estudis Espacials de Catalunya (IEEC) 08034 Barcelona, Catalonia, Spain}

\author[0000-0002-9637-4554]{Albert Rimola}
\affiliation{Departament de Química, Universitat Autònoma de Barcelona 08193 Bellaterra, Catalonia, Spain}



\begin{abstract}

NASA's DART and ESA's Hera missions offer a unique opportunity to investigate the delivery of impact ejecta to other celestial bodies. We performed ejecta dynamical simulations using 3 million particles categorized into three size populations (10 cm, 0.5 cm, and 30 $\mu$m) and constrained by early post-impact LICIACube observations. The main simulation explored ejecta velocities ranging from 1 to 1,000 m/s, while a secondary simulation focused on faster ejecta with velocities from 1 to 2 km/s. We identified DART ejecta orbits compatible with the delivery of meteor-producing particles to Mars and Earth. Our results indicate the possibility of ejecta reaching the Mars Hill sphere in 13 years for launch velocities around 450 m/s, which is within the observed range. Some ejecta particles launched at 770 m/s could reach Mars's vicinity in 7 years. Faster ejecta resulted in a higher flux delivery towards Mars and particles impacting the Earth Hill sphere above 1.5 km/s. The delivery process is slightly sensitive to the initial observed cone range and driven by synodic periods. The launch locations for material delivery to Mars were predominantly northern the DART impact site, while they displayed a southwestern tendency for the Earth-Moon system. Larger particles exhibit a marginally greater likelihood of reaching Mars, while smaller particles favor delivery to Earth-Moon, although this effect is insignificant. To support observational campaigns for DART-created meteors, we provide comprehensive information on the encounter characteristics (orbital elements and radiants) and quantify the orbital decoherence degree of the released meteoroids.

\end{abstract}
\keywords{Impact phenomena (779), Ejecta (453), Meteors (1041), Orbital evolution (1178)}


\section{Introduction} \label{sec:intro}
The study of asteroid impacts and their effects on celestial bodies has garnered significant attention in recent years due to the potential threat they pose to Earth and the need to develop effective planetary defense strategies. NASA's Double Asteroid Redirection Test (DART) \citep{Cheng2018PSS, Rivkin2021PSJ} and the European Space Agency's (ESA) Hera mission \citep{Michel2022PSJ} are the first planetary defense missions. They aim to test the feasibility of using a kinetic impactor to divert a potentially hazardous asteroid from a collision course with Earth. The main goal of DART is to demonstrate and validate the deflection technology, while Hera investigates the outcome of the impact. By studying the results and understanding the dynamics of the impact, valuable insights can be gained to enhance our capabilities in deflecting other asteroids in the future.




Launched in 2021, the DART mission was specifically designed to impact Dimorphos, the moon of the binary asteroid system (65803) Didymos, to modify its orbit. The DART impact occurred on September 26, 2022, at a velocity of 6.14 km/s in a virtually head-on manner, reducing the orbital period of Dimorphos around Didymos by -33.25 minutes with an error of a few seconds \citep{Scheirich2024PSJ, Naidu2024PSJ}, most likely changing Dimorphos’ spin state into non-principle axis rotation \citep{Pravec2024Icar}, and producing ejecta that were observed from multiple space- and ground-based observatories \citep{Cheng2023Natur, Daly2023Natur, Li2023Natur, Thomas2023Natur}. DART included the Light Italian CubeSat for Imaging of Asteroid (LICIACube) which was hosted as a piggyback and released two weeks before the impact and observed the results of the collision during a flyby of the Didymos system. LICIACube captured valuable data on the collision and monitored the initial aftermath for a brief period of time, offering a first constraint on the ejecta properties \citep{Dotto2021PSS}. 

Hera is scheduled to be launched in October 2024 and is expected to rendezvous with Didymos in late 2026. The primary goals of the Hera mission, along with its two cubesats, Juventas \citep{Goldberg2019} and Milani \citet{Ferrari2021AdSpRMilani}, are threefold. First, it aims to accurately measure the mass of Dimorphos to gain insights into the efficiency of momentum transfer resulting from the DART impact. Second, it seeks to conduct a detailed study of the impact crater or the global reshaping of Dimorphos, thereby enhancing our understanding of the collision process. Lastly, Hera aims to investigate the properties of the target asteroids to facilitate the scaling of the impact outcome to other potential targets.

Concerning the ejecta, Hera's primary objective is to investigate the presence of newly deposited material on Didymos and Dimorphos, which may include reaccreted ejecta resulting from the DART impact. To accomplish this, Hera will utilize its onboard cameras, such as the Asteroid Framing Cameras on Hera and the navigation cameras on Milani and Juventas, to observe albedo variations. Additionally, its visible and near-infrared spectrometers, namely Hyperscout on Hera and ASPECT on Milani, will be employed to search for spectral signatures indicative of fresh ejecta material.

The impact of DART on Dimorphos and the subsequent ejection of debris present an opportunity to examine various aspects related to the dynamics and fates of the ejecta. We investigate the potential delivery of meteoroid-sized ($\mu$m to cm scale) materials from Dimorphos to other celestial bodies using a \textit{N}-body numerical integrator. Such simulation is valuable for understanding the dynamical evolution of the launched fragments and planning DART-created meteor observation campaigns, similar to the work conducted by \citet{Jenniskens2023Icar} for Bennu's meteoroids. 


Preliminary investigations have been conducted to examine both the short-term dynamics of the ejecta \citep{Larson2021MNRAS, Nakazawa2021PSJ, Ferrari2022PSJ, ferrari2024morphology} and longer-term evolution \citep{Yu2017Icar, Yu2018Icar, Rossi2022PSJ, Fenucci2024MNRAS5286660F}. Specifically, the delivery of particles from Dimorphos to Earth through the DART impact was modeled by \citep{Wiegert2020PSJ}. However, none of these studies have utilized post-impact observational data and tracked the particles to assess their potential impacts on other bodies beyond the binary system.

In this paper, we aim to further our knowledge of the mid-term dynamical evolution of the ejecta resulting from the DART impact on Dimorphos, specifically focusing on particles exceeding the system's escape velocity. Our analysis considers multiple post-impact observations, incorporating the influence of solar radiation forces. By tracking the trajectories of the ejected particles, we investigate the potential delivery to planetary bodies within the solar system. Our research emphasizes the delivery of material toward Mars or Earth. 

Our work exemplifies how novel techniques enable the investigation of the evolutionary processes of small bodies in the solar system, pushing early forecasting of impact hazards. Missions such as DART and Hera offer a valuable opportunity to engage the public in asteroid studies and explore scientific outreach and popularization avenues. Planetary defense is a compelling example of applied multidisciplinary science, benefiting humanity \citep{Trigo2022}.

We provide a detailed account of the applied methodology and simulation conditions in Section \ref{sec:method}, present a comprehensive analysis of the results in Section \ref{sec:results}, and summarize the major findings in Section \ref{sec:conclusions}.

\section{Methods and simulation set-up} \label{sec:method}

To predict the fate of the particles ejected by the DART impact on Dimorphos, we calculate their trajectories under the influence of the gravity of Didymos and Dimorphos, the Sun and Mercury, Venus, Earth, Mars, and Jupiter. We also include the Moon as we account for the Earth-Moon barycenter. To accurately model the main ejecta, it is essential to consider the non-spherical gravitational effects near the surfaces of Dimorphos and Didymos. However, for the fast ejecta well above the system's escape velocity, this consideration is unnecessary because particles at that speed do not feel the influence of the binary system. Therefore, modeling the mass as a point source is sufficient for our study.

For a particle with mass $m$ and geometrical cross section $A$, moving at velocity $\mathbf{v}$ through a radiation field with energy flux density $S$, the equation of motion to terms of order $v/c$ is given by:

\begin{equation}\label{eq:radiation}
    m\dot{\mathbf{v}} = \left( \frac{S A}{c} \right) Q_{\text{pr}} \left[ \left( 1 - \frac{\dot{r}}{c} \right) \hat{S} - \frac{\mathbf{v}}{c} \right]
\end{equation}

Here, $\hat{S}$ is a unit vector pointing in the direction of the incoming radiation, $\dot{r}$ represents the particle's radial velocity, and $c$ is the speed of light. The radiation pressure efficiency factor $Q_{\text{pr}}$ is assumed to be one when the grain size exceeds the wavelength of the radiation.

A radius is assigned randomly to each particle in the typical range of meteoroid size producing observable meteoroids, with options being 10 cm, 0.5 cm, or 30 $\mu$m. Conventional optical meteor systems usually detect millimeter-sized particles, with specialized sensors capable of detecting down to a limiting magnitude of 12 \citep{Watanabe2014me13conf}, equivalent to micrometer diameters. We set the lower size limit at 30 $\mu$m, aligning with the lower bound of the meteoroid consensus definition, as this is the smallest size detectable by radars \citep{Janches2015ApJ, 2017JIMO...45...91K}. Larger particles will undergo similar dynamic evolution, as radiation effects are negligible for these sizes within the timescales under consideration. We assume a grain density of 3,350 $kg,m^{-3}$, which is within the mid-range of values typically used in DART ejecta studies \citep{Raducan2024NatAs}, as reflectance spectra analysis of Didymos suggests that the most suitable meteorite analogs for Dimorphos' boulders are L/LL ordinary chondrites, which exhibit grain densities ranging from approximately 3,200 to 3,500 kg m$^{-3}$ \citep{Flynn2018ChEG}. We assume a mass of 4.3$\cdot$10$^9$ kg for Dimorphos and 5.557$\cdot$10$^{11}$ kg for Didymos, and set an ellipsoidal shape of 177$\times$174$\times$116 m and 851x849$\times$620 m respectively \citep{Daly2023Natur}.

Based on LICIACube observations, the ejecta resulting from the impact displays an elliptic cone shape \citep{Deshapriya2023PSJ}. The cone's axis is directed towards R.A., Dec. (in J2000): $147^{+1^\circ}_{-10^\circ}$, $+16^{+4^\circ}_{-6^\circ}$. The cone is described by two perpendicular half-angles: $\eta = 69^{+1^\circ}_{-3^\circ}$ and $\gamma = 51^{+1^\circ}_{-11^\circ}$, with a rotational angle of $\omega = 12^\circ$ around its axis. The intersection of the observed cone and the surface of Dimorphos corresponds to a crater with a maximum radius of 65 meters \citep{Deshapriya2023PSJ}, which is the value we adopt here. The cone's apex is positioned at the impact site on Dimorphos' surface. To test the cone dependency, as initial conditions in our simulation, the particles are ejected from a crater centered on the DART impact site at coordinates 8.84$^\circ$ S, 264.30$^\circ$ E in the Dimorphos body-fixed frame \citep{Daly2023Natur} in three configurations: with the nominal values of the observed elliptical cone (nominal), with its 1-$\sigma$ upper bound (upper), and with its 1-$\sigma$ lower bound (lower).

Using images acquired with two instruments—the LICIACube Explorer Imaging for Asteroid (LEIA) and the LICIACube Unit Key Explorer (LUKE)—ejecta velocities were measured to range from a few tens of meters per second to about 500 meters per second \citep{Dotto2024Natur}. We perform a main simulation with 3 million particles with ejecta velocities of 1 to 1000 m/s to cover this observed range, where the lower limit corresponds to the asteroid system escape velocity \citep{Michel2016AdSpR}. We emphasize in our results the particles ejected at less than 500 m/s. \citet{Ofek2023} utilized observations from several instruments to study the DART impact. Data were collected using the Large Array Survey Telescope (LAST), the Swift-UVOT space telescope, and the Jay Naum Rich 28-inch telescope (C28) at the Wise Observatory. These observations revealed a mean ejecta velocity of 1.6$\pm$0.2 km/s for sub-micrometer particles. In any case, we conduct a secondary simulation with 1 million particles to account for these higher velocities, ranging from 1 to 2 km/s. 

We employ the \texttt{REBOUND} package \citep{Rein2015MNRAS, Rein2017MNRAS} and its extension module \texttt{REBOUNDx} package \citep{Tamayo2020MNRAS}. It is a popular open-source \texttt{Python} code used for simulating N-body orbital dynamics that provides a flexible framework for integrating the motion of particles under the influence of gravitational and other forces. Among the various options the package offers, we employ the Gragg-Bulirsch-Stoer integrator. This powerful numerical integration method combines the Richardson polynomial extrapolation and the modified midpoint method to solve ordinary differential equations. One of its key features is adaptivity, as it dynamically adjusts the step size during the integration process. This ensures accurate results while optimizing computational resources, particularly in scenarios involving close encounters \citep{hairer1993solving}. We add the radiation force using \texttt{REBOUNDx} based on Eq. \ref{eq:radiation}.

We perform a numerical integration of the system over 100 years. To improve our statistics, we track the particles that reach the Hill sphere of a planetary body. This approach is necessary because a much larger number of particles would be required to find direct impacts with an object. For each particle impacting the Hill sphere, we record its position and velocity vector. We then perform a statistical estimation to determine the number of particles in the simulation required to obtain direct impacts. Specifically, we calculate the distribution of perpendicular distances of the particle velocity vectors to the planetary body, fit a multivariate density distribution function \citep{Scott2015mdetboook}, and generate new distances until some are smaller than the effective cross-section of the planet, considering gravitational focusing.

Table \ref{tab:parameters} shows the general parameters for the simulation. The ephemerides are obtained from the kernel $dart\_v01.tm$ via the WebGeocalc toolkit that NASA's Navigation and Ancillary Information Facility (NAIF) provides \citep{Acton1996PSS, Acton2018PSS}. Note that in the secondary simulation (fast-rnd or also denoted with $^*$ in the figure), the one with the faster ejecta, we investigate the influence of the launch position within the crater by assigning random velocities to the particles. Unlike the main simulation, where particle velocities decrease radially with increasing distance from the impact site, this approach allows us to assess the effects of varied launch locations and velocities while maintaining the cone direction.

\begin{table}[ht]
	\centering
	\caption{General parameters of the two simulations.}
	\label{tab:parameters}
    \begin{tabular}{lc}
    	\hline
        Parameter & Value                           \\
        \hline
        Simulation & Main $\mid$ Fast-rnd \\
        Duration & 100 years \\
        Time step & Adaptive \\
        No. particles & 3$\cdot$10$^6$ $\mid$ 1$\cdot$10$^6$ \\
        Impact location & 8.84$^\circ$ S, 264.30$^\circ$ E \\
        Particle radius & 10 cm, 0.5 cm, 30 $\mu$m \\
        Particle density & 3,350 $kg\,m^{-3}$ \\
        Ejecta velocity & 1 -- 1,000 m/s $\mid$ 1 -- 2 km/s \\
        Ejecta vel-pos & Radial $\mid$ Random \\
        Cone direction & R.A.=$147^{+1^\circ}_{-10^\circ}$, Dec.=$+16^{+4^\circ}_{-6^\circ}$ \\
        Cone aperture & $\eta$=$69^{+1^\circ}_{-3^\circ}$, $\gamma$=$51^{+1^\circ}_{-11^\circ}$\\
        Cone rotation & $12^\circ$\\
        Dimorphos mass & 4.3$\cdot$10$^9$ kg \\
        Dimorphos ellipsoid & 177$\times$174$\times$116 m \\
        Didymos mass & 5.557$\cdot$10$^{11}$ kg \\
        Didymos ellipsoid & 851$\times$849$\times$620 m\\
        Sun luminosity & 3.85$\cdot$10$^{26}$ W \\
        \hline
    \end{tabular}
\end{table}

Finally, we assess the potential DART-created meteor showers observable on both Mars and Earth by evaluating the expected values for traditional orbital similarity criteria. Orbital elements such as inclination $i$, eccentricity $e$, longitude of the ascending node $\Omega$, perihelion distance $q$, and argument of perihelion $\omega$ define the path of any object following a Keplerian trajectory. By comparing these elements, we can identify connections between any two orbits, subject to a specific threshold. The first method, known as D-criteria, was developed in the latter half of the 20th century by \citet{Southworth1963SCoA....7..261S}:

\begin{multline}
D_{SH}^{2}=\left(e_{B}-e_{A}\right)^{2}+\left(q_{B}-q_{A}\right)^{2}+\left(2 \sin \frac{I_{AB}}{2}\right)^{2}+ \\
+\left(\frac{e_{B}+e_{A}}{2}\right)^{2}\left(2 \sin \frac{\pi_{B A}}{2}\right)^{2},
\label{eq
}
\end{multline}

where $\pi_{BA}$ is the angle between the perihelion points, and $I_{AB}$ is the angle between the inclinations of the orbits.

\citet{Drummond1981Icar...45..545D} refined this approach by introducing the angle between the perihelion points $\theta_{BA}$ and by weighting the terms $e$ and $q$ to ensure each term contributes equally to the overall metric. This variant, known as $D_{D}$, is defined as:

\begin{multline}
D_{D}^{2}=\left(\frac{e_{B}-e_{A}}{e_{B}+e_{A}}\right)^{2}+\left(\frac{q_{B}-q_{A}}{q_{B}+q_{A}}\right)^{2}+\left(\frac{I_{AB}}{\pi}\right)^{2} + \\
+\left(\frac{e_{B}+e_{A}}{2}\right)^{2}\left(\frac{\theta_{B A}}{\pi}\right)^{2}.
\label{eq
}
\end{multline}

Both criteria are widely used today to associate a meteor with a meteoroid stream or a parent body \citep{Wiegert2004EMP, Trigo2007MNRAS, Jopek2013MNRAS, Dumitru2017AA, Jenniskens2017PSS, Vaubaillon2019, Eloy2021MNRAS, Eloy2022AJ, Eloy2023MNRAS, PeaAsensio2024}. We then compute $D_{SH}$ and $D_{D}$ first to quantify the degree of decoherence of the released meteoroids and second to offer a practical tool for future DART-created meteor campaigns.


\section{Results} \label{sec:results}

\begin{figure*}
\includegraphics[width=\linewidth]{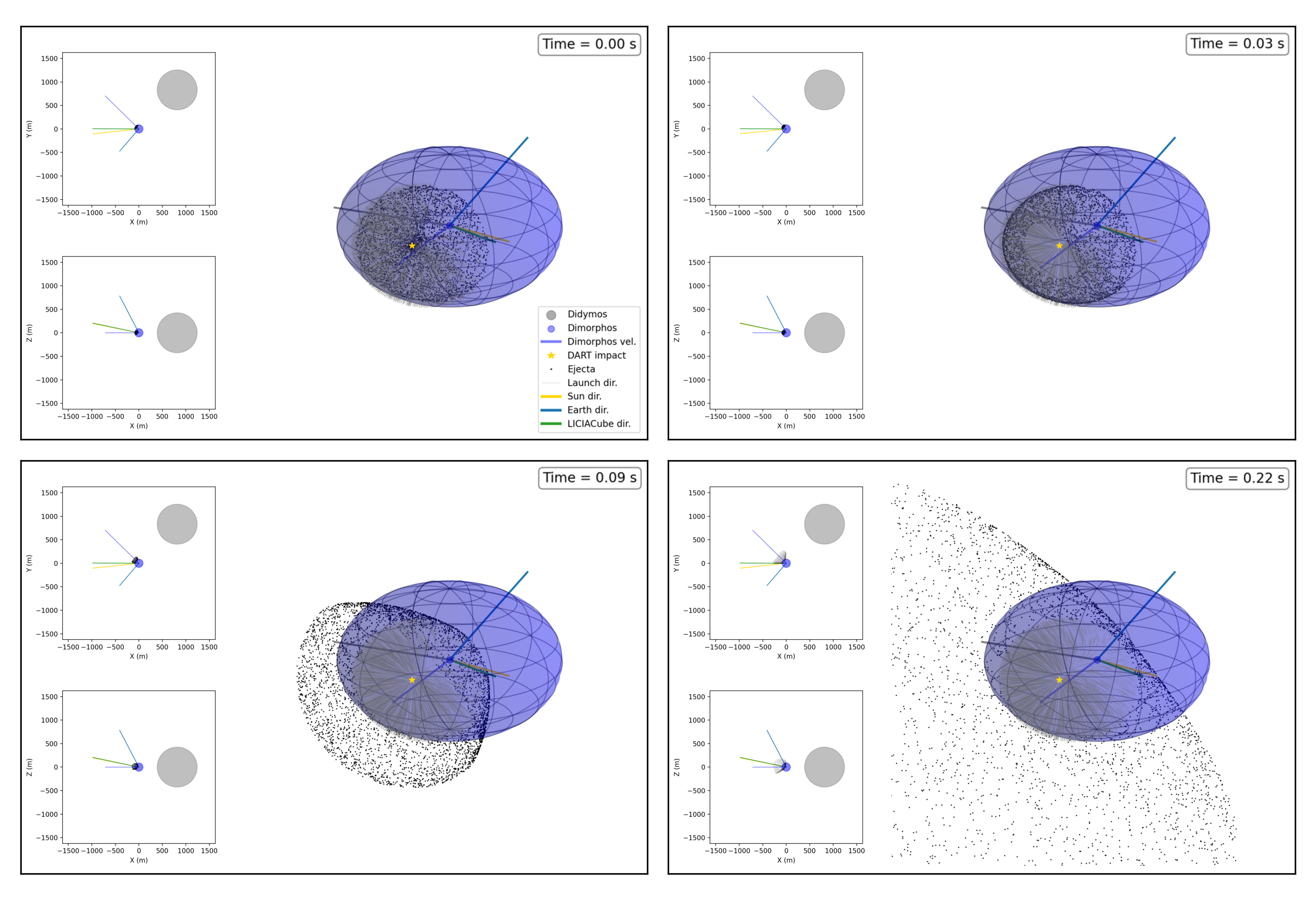}
\caption{Early snapshots of the ejecta evolution centered on Dimorphos. The four panels show the progression of the ejecta at different time intervals. Vectors are included to illustrate the geometry of the initial conditions, as well as X-Y and X-Z projected views.
\label{fig:ejecta_config}}
\end{figure*}

\begin{figure}[h!]
\centering
\includegraphics[width=\linewidth]{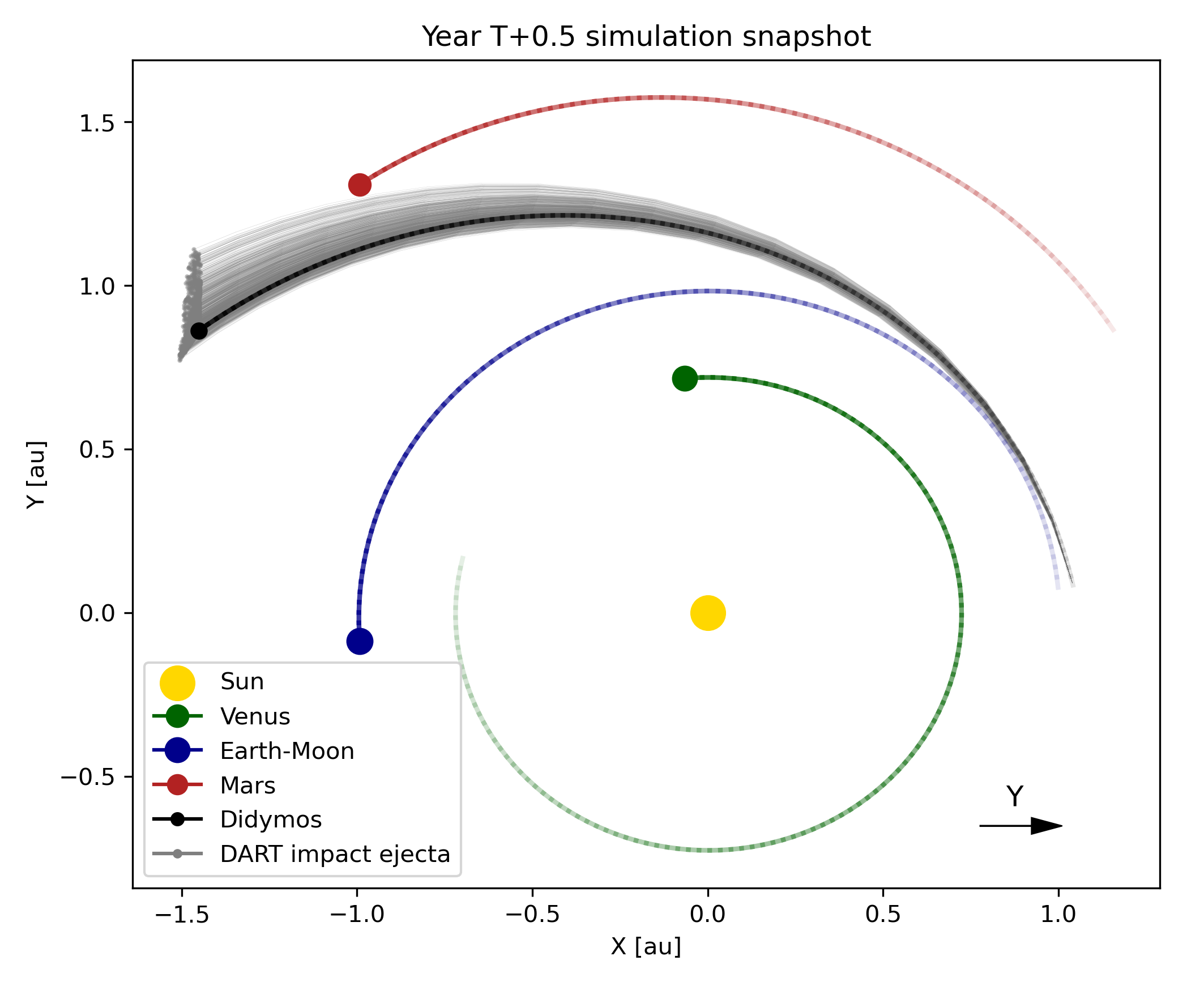}
\caption{Illustrative snapshots of the main simulation a 0.5 years post-impact.
\label{fig:orbits_eclip}} 
\end{figure}

Figure \ref{fig:ejecta_config} presents four early snapshots of the ejecta evolution centered on Dimorphos within the first quarter of a second. The figure includes vectors that define the geometry of the initial conditions, as well as X-Y and X-Z projected views. The fate statistics of our 100-year simulations are summarized in Table \ref{tab:stats_results}. The table provides the percentage of particles of various sizes that reach the Hill spheres of Mars and the Earth-Moon system, along with their minimum ejection velocities, the time taken for the delivery, the earliest delivery and slowest ejection velocities, and the corresponding delivery times. Figure \ref{fig:orbits_eclip} provides an illustrative snapshot of the simulation at 0.5 years post-impact. The early divergence of the ejecta after the impact is evident, crossing Mars' orbit before intersecting with it.

\begin{table*}
\normalsize
\caption{Ejecta fate statistics for 100-year simulation. The table presents the percentage of ejecta particles (10 cm, 0.5 cm, and 30 $\mu$m) reaching the Hill spheres of Mars and the Earth-Moon system, along with their minimum ejection velocities ($v_{ej,min}$), delivery times corresponding to these velocities, the earliest delivery and slowest ejection velocities ($v_{e,s}$), and delivery times for the earliest delivery and slowest ejection velocities. The main simulation is grouped by the different cone configurations (upper, nominal, lower). General values are given for the secondary simulation (fast-rnd). ipm: Mars/Earth direct impacts per million of simulated particles.}\label{tab:stats_results}
\begin{tabular}{lccccccccc}
\hline
& 10 cm & 0.5 cm & 30 $\mu$m & $v_{ej,min}$ & Time $v_{ej,min}$ & $v_{e,s}$ & Time $v_{e,s}$ & ipm \\
& [\%] & [\%] & [\%] & [m/s] & [year] & [m/s] & [year] &  \\
\hline
\multicolumn{9}{c}{\textbf{To Mars Hill sphere}} \\
Upper & 0.181 &	0.182 & 0.151 & 455.382 & 12.747 & 782.478 & 7.117 & 1.691 \\
Nominal & 0.194 & 0.199 & 0.178 & 449.201 & 27.779 & 772.765 & 7.117 & 1.222 \\
Lower & 0.271 & 0.276 & 0.196 & 438.610 & 14.624 & 781.505 & 7.116 & 1.481 \\
Fast-rnd & 0.370 & 0.358 & 0.357 & 1000.133 & 61.609 & 1217.262 & 5.244 & 2.119 \\

\multicolumn{9}{c}{\textbf{To Earth-Moon Hill sphere}} \\
Upper & 0 & 0 & 0 & 0 & 0 & 0 & 0 & 0 \\
Nominal & 0 & 0 & 0 & 0 & 0 & 0 & 0 & 0 \\
Lower & 0 & 0 & 0 & 0 & 0 & 0 & 0 & 0 \\
Fast-rnd & 0.248 & 0.252 & 0.231 & 1518.371 & 32.103 & 1807.255 & 7.103 & 1.015 \\

\hline
\end{tabular}
\end{table*}

The ipm (impacts per million) values presented in Tabel \ref{tab:stats_results} indicate the number of direct impacts with Mars or Earth per million of particles simulated, so it represents the number of direct impacts that can be expected in the simulation. Based on these numbers, we would expect approximately at least four direct impacts with Mars in our main simulation and around one impact with the Earth in the fast-rnd simulation. The ipm values provide an indication of how many simulated particles would need to be increased for specific initial conditions to achieve a certain number of direct impacts.

\begin{figure*}
\centering
\includegraphics[width=0.9\linewidth]{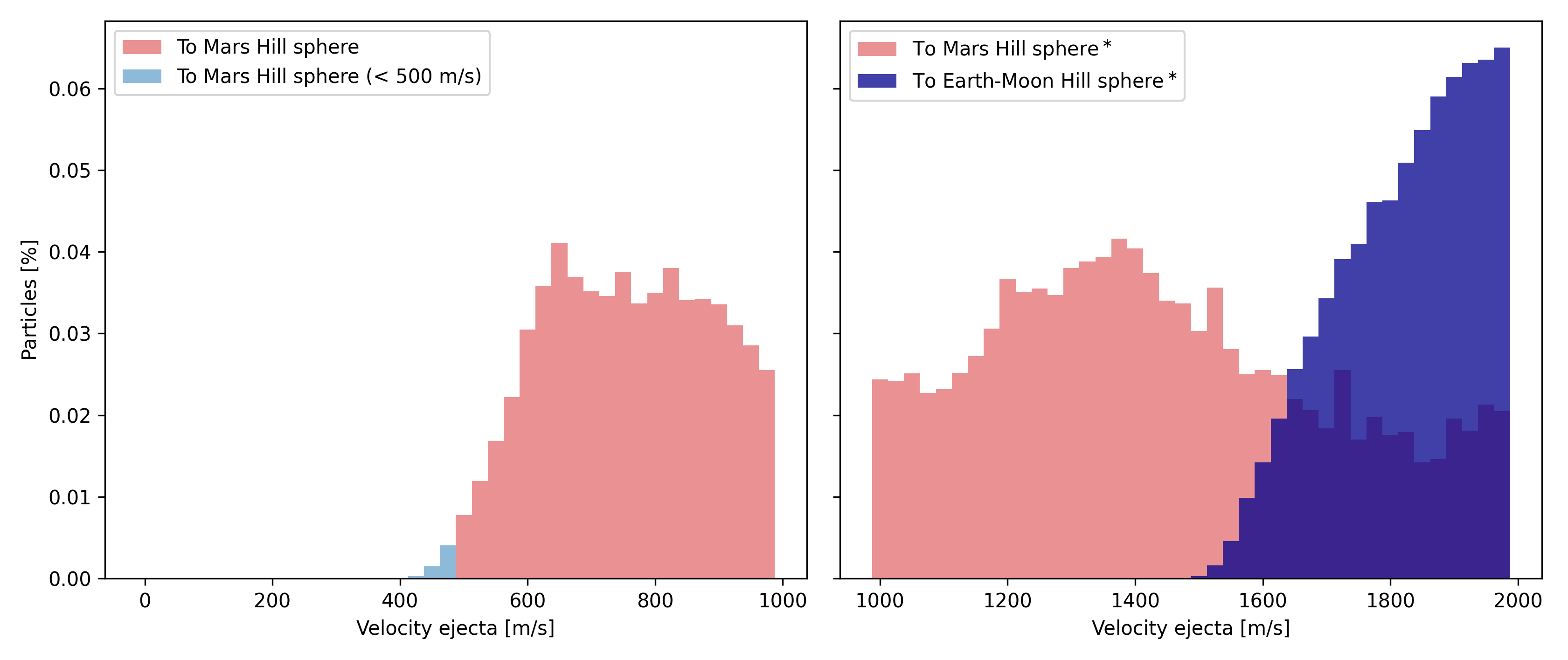}
\caption{Distribution of ejecta velocities and their delivery outcomes. The left panel shows the percentage of particles reaching the Mars Hill sphere, with a focus on particle ejecta velocities at velocities below 500 m/s. The right panel displays the ejecta velocities distribution of delivered particles for the fast-rnd denoted by a $^*$.}
\label{fig:vel_ejec}
\end{figure*}

The delivery process exhibits slight sensitivity to the initial observed cone range, with the percentage of particles reaching the Hill spheres being similar across the configurations\footnote{We also tested small variations in cone direction and aperture angles and found that, while the total delivery remains constant, the minimum delivery time varies, with some configurations achieving delivery in as early as 2 years.}. The upper bound cone configuration shows a higher reduction in the delivery flux, as opposed to the lower case. The results highlight the potential for material from the DART impact to reach the Mars Hill sphere within 13 years for launch velocities of $\sim$450 m/s, which falls within the observed range of ejection velocities. Ejecta particles launched at 770 m/s could reach the vicinity of Mars in approximately 7 years. Additionally, simulations of faster ejecta showed an increased flux delivery towards Mars, with the first deliveries in 5 years. For the Earth-Moon Hill sphere, particles with velocities above 1.5 km/s can reach the vicinity within $\sim$7 years. Figure \ref{fig:vel_ejec} shows the ejecta velocity histogram for both simulations. A delivery peak to Mars is observed at 625 m/s and 1.4 km/s, whereas the number of particles delivered to the Earth-Moon system continues to increase with higher launch velocities. Note that the orbit of Didymos crosses Mars' orbit but remains outside Earth's orbit (q $<$ 1.013 au). Consequently, achieving a trajectory that brings the ejecta into proximity with Earth requires a higher velocity compared to intersecting Mars' orbit. Regarding particle size, centimeter-sized particles exhibit a similar percentage of delivery. In contrast, micrometer-sized particles show a reduction, indicating that radiation forces pushing them into farther orbits hinder the delivery.

\begin{figure*}
\centering
\includegraphics[width=1\linewidth]{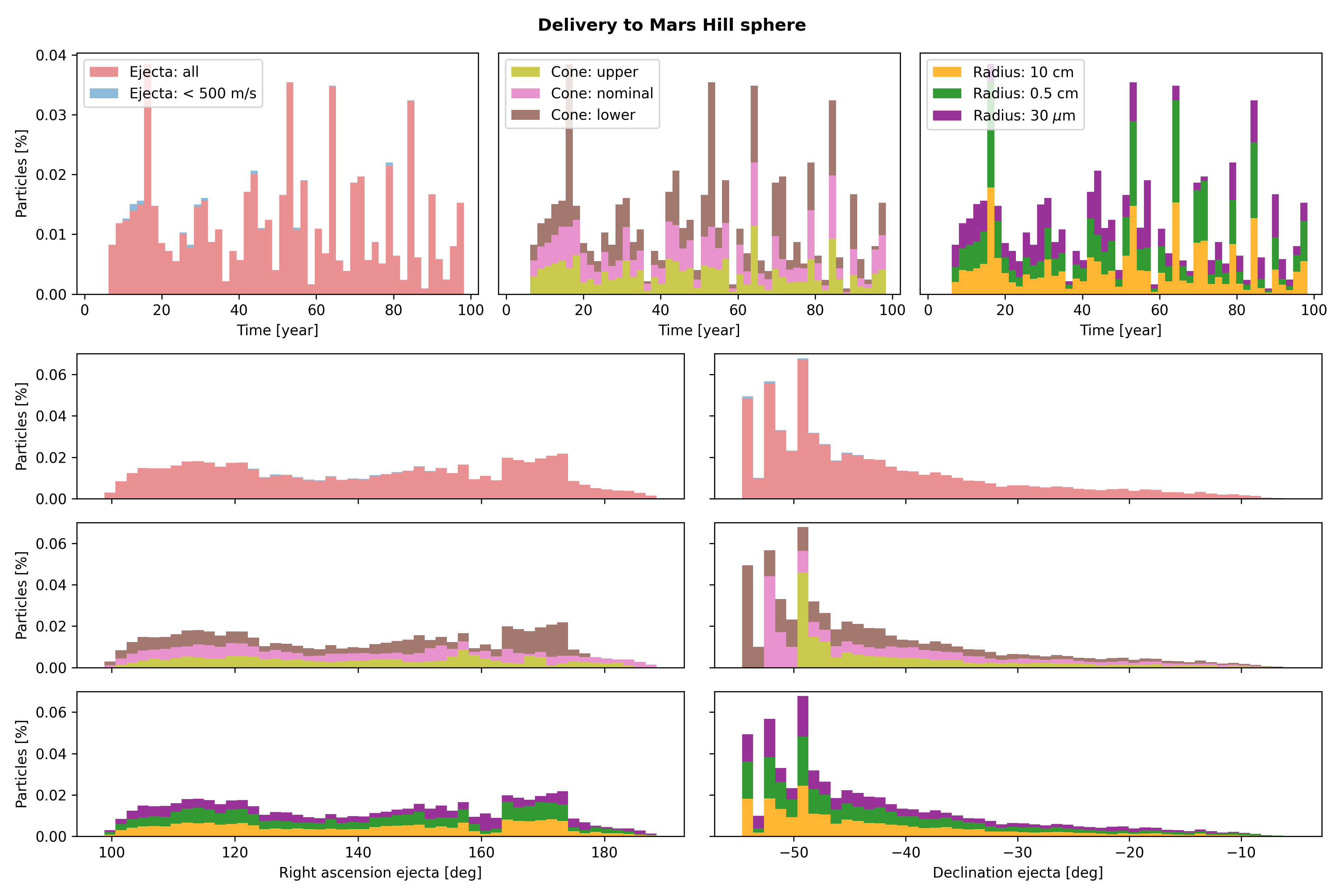}
\caption{The first row shows the histograms of delivery time to the Mars Hill sphere. The next three rows display the histograms of the ejecta direction in the J2000 reference frame centered on Dimorphos that produce delivery to the Mars Hill sphere. The plots are grouped below or above 500 m/s of launch velocity, by the cone configurations, and by particle radius. All plots refer to the main simulation results.}
\label{fig:porc_imp_time_ra_dec_hist}
\end{figure*}

Figure \ref{fig:porc_imp_time_ra_dec_hist} provides an overview of the initial direction dependency of the ejecta delivered to the Mars Hill sphere. The first row of histograms illustrates the delivery time of these particles. There is no significant difference in the delivery times or the amount of particles delivered across different initial cone configurations. It is possible to observe some peaks of augmented delivery flux. The subsequent three rows display histograms of the ejecta direction in the J2000 reference frame centered on Dimorphos. While the amount and delivery times of particles remain consistent across different cone configurations and particle sizes, distinct patterns emerge in the directionality of the ejecta. An almost uniform distribution is observed in the right ascension, except for the lower cone configuration, which exhibits a concentration of delivery around 170$^\circ$. In declination, all cases show an accumulation trend over -50$^\circ$.

\begin{figure}[H]
\centering
\includegraphics[width=1\linewidth]{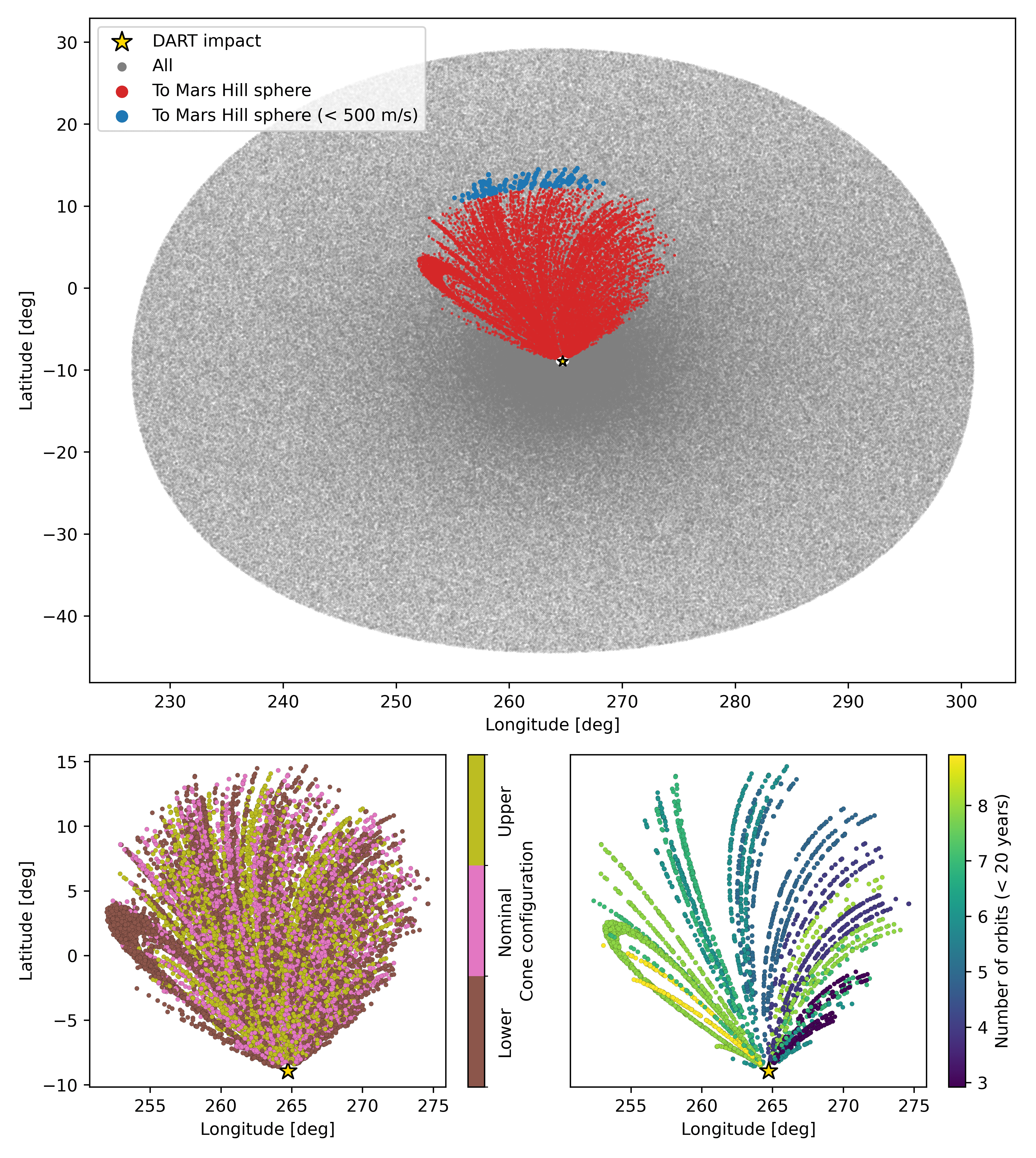}
\caption{Ejecta launch locations of particles delivered to Mars in the Dimorphos body-fixed frame for the main simulation group by particles that were launched above and below 500 m/s (top), by cone configuration (bottom left), and by the number of orbit pre-impact for the first 20 years (bottom right).}
\label{fig:location_eject_radiant}
\end{figure}

The ejecta launch locations represented in the Dimorphos body-fixed frame demonstrate a confined region of potential material delivery to Mars, as depicted in Figure \ref{fig:location_eject_radiant}. The top panel illustrates the initial positions of particles ejected at velocities below 500 m/s. Note that our defined distribution of velocities based on radial distances from the DART impact prevents them from being any closer or further. As we will see later, the important initial conditions are the direction and velocity, not the position. The bottom left panel shows the delivered particle positions relative to each cone configuration, revealing a distinguishable pattern, although equally distributed. Various combinations of launch parameters (position, direction, and velocity) result in delivery within the same time window. These combination patterns are even more accentuated in the bottom right panel, which depicts the number of orbits pre-impact for the first 20 years. This implies that the particles impacting the Mars Hill sphere are further subclustered by the number of orbits they have completed around the Sun, indicating that the delivery is driven by synodic periods. A characteristic structure is observed at lower longitudes and latitudes, which represents an efficient delivery channel that appears exclusively in the lower cone configuration and coincides with the right ascension peak over 170$^\circ$ (Fig. \ref{fig:porc_imp_time_ra_dec_hist}).

\begin{figure}
\centering
\includegraphics[width=1\linewidth]{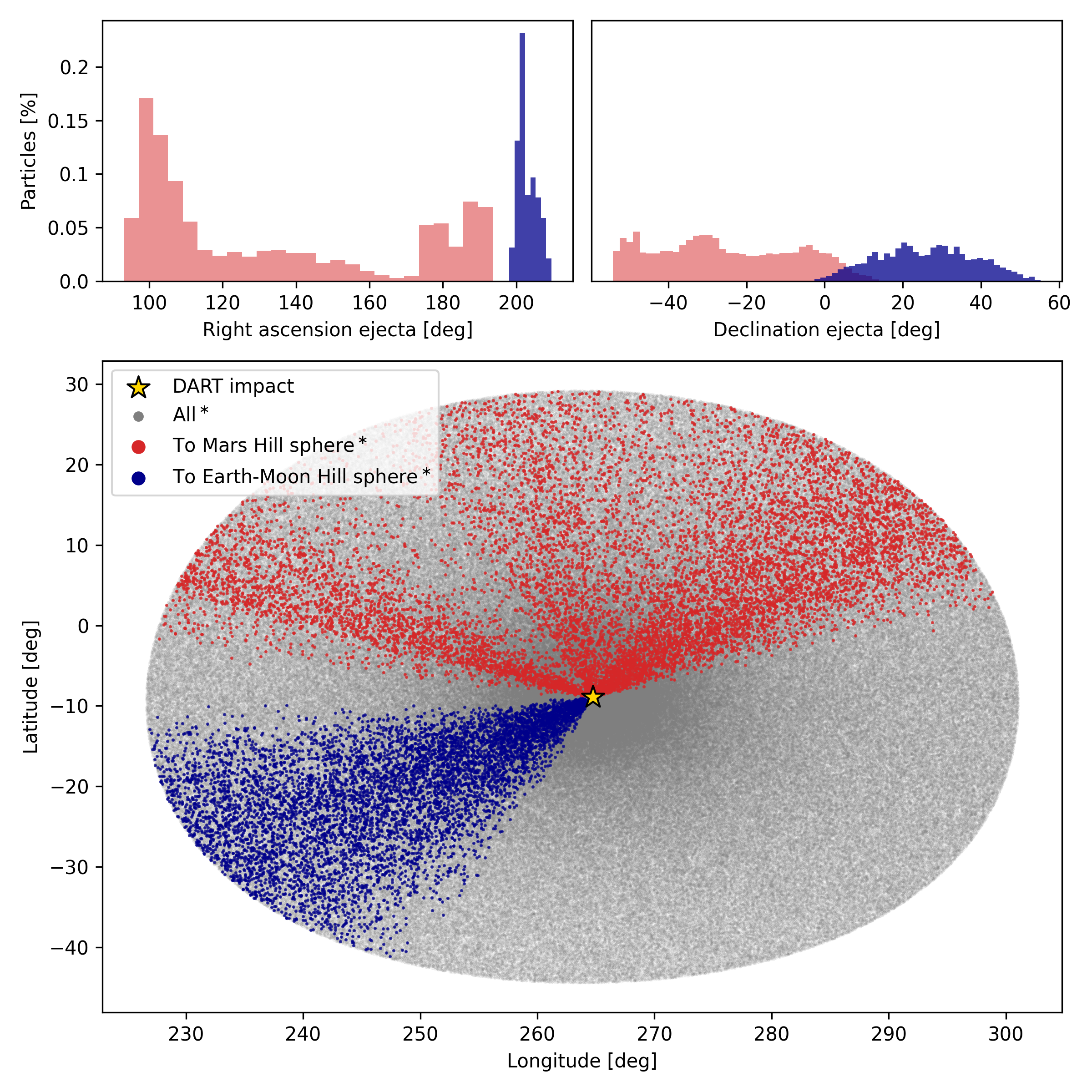}
\caption{Ejecta direction in the J2000 reference frame centered on Dimorphos that produce delivery to the Mars and Earth-Moon Hill spheres for the fast-rnd simulation, top left showing the right ascension and top right the declination. Top bottom panel displays the ejecta launch locations of delivered particles in the Dimorphos body-fixed frame for the fast-rnd simulation. The $^*$ denotes the fast-rnd simulation for clarity.}
\label{fig:location_eject_radiant_fast}
\end{figure}

Regarding the fast-rnd simulation, the launch positions that result in material delivery with very high launch velocities to Mars and the Earth-Moon system are clearly defined and distinct, as shown in Figure \ref{fig:location_eject_radiant_fast}. Compared to the main simulation, the region of material that could reach Mars has expanded, occupying nearly the entire northern area of the DART impact site (bottom panel). On the other hand, the region for delivery to the Earth-Moon Hill sphere is more confined and biased towards the southwest of the impact. Compared to the main simulation, the Mars delivery expanded the range of ejecta direction in both right ascension and declination. In declination, it partially overlaps with the Earth-Moon delivery at its minimum, whereas in right ascension, the ranges do not overlap, with the Earth-Moon delivery being remarkably concentrated. Note that delivery occurs regardless of the launch location.

To facilitate the observation campaigns of potential DART-created meteors on Mars, we provide the encounter characteristics of the potential delivered material to Mars and the Earth-Moon. We have compiled all the key impact parameters for all simulations, grouped by initial conditions to delivery time. Table \ref{tab:compilation} presents the right ascension and declination of the ejecta direction in the J2000 frame, its velocity, launch position in the Dimorphos body-fixed frame, and the right ascension and declination of the radiant (the particle anti-apex) in J2000 frame, the relative velocity, and the orbital elements in ECLIPJ2000 at the time of impact with the respective Hill spheres. Both the semi-major axis and eccentricity of the delivered particles are larger than the initial values in the Didymos system (1.643 au and 0.383, respectively). This can be attributed to the fact that the DART impact was nearly opposite to the relative motion of Dimorphos around Didymos and nearly opposite to the binary system's motion around the Sun. Consequently, the ejecta from the impact were launched with higher heliocentric velocity and energy, increasing both the eccentricity and the semi-major axis of the particles. It is also noteworthy that the delivered particles tend to have the lowest inclinations, as these favor encounters due to the coincidence of the synodic period.

Figure \ref{fig:fate_radiant} of the Appendix showcases the right ascension and declination of the radiant in the J2000 reference frame centered on Mars, as well as the relative impact velocity and launch velocity in relation to Mars' solar longitude and delivery time to the Mars Hill sphere. This data is presented for particles launched at velocities below 500 m/s, and categorized by the cone configuration and different sizes. Figure \ref{fig:fate_orbelem} displays the orbital elements in the ECLIPJ2000 reference frame at impact with the Mars Hill sphere, subdivided into the same three categories for Mars' solar longitude and delivery time. Similarly, Figures \ref{fig:earth_fate_radiant} and \ref{fig:earth_fate_orbelem} of the Appendix show the equivalent for the delivery to Earth of the fast-rnd simulation. We note that the impact velocities are relative to the Hill spheres (similar to the so-called geocentric velocities), so a direct impact with Mars and Earth should account for their respective gravitational accelerations. The clustering observed in the semi-major axis, eccentricity, and inclination further suggests that the timing of impacts is synchronized with the relative positions to Mars and the Earth, leading to periodic clustering of impacts based on their orbital dynamics. Both the radiant and the relative impact velocity exhibit well-defined clusters. The dispersion of the clustering reduces as the number of years considered decreases. The radiant position indicates that DART-created meteors in the Earth's atmosphere will be predominantly observable from the southern hemisphere, although lower latitudes in the northern hemisphere may also witness grazing meteors. The solar longitude and the orbital elements at impact also exhibit clear clustering, reinforcing the idea of the delivery being driven by synodic periods.

\begin{sidewaystable}
\tiny
	\centering
	\caption{Compilation of average initial (eje.) and impact (imp.) parameters and their standard deviations for the material delivered to Mars and the Earth-Moon system (E-M) Hill sphere. Impact parameters are given in J200, while the orbital elements are in ECLIPJ2000. The fast-rnd simulation is denoted by $^*$.}
	\label{tab:compilation}
    \begin{tabular}{lccccccccccccccc}
    	\hline
        Fate & Group & Sol. Lon. & R. A. (eje.) & Dec. (eje.) & Vel. (eje.) & Lat. (eje.) & Lon. (eje.) & R. A. (imp.) & Dec. (imp.) & Vel. (imp.) & a & e & i & $\omega$ & $\Omega$    \\
          &   & [deg] & [deg] & [deg] & [m/s] & [deg] & [deg] & [deg] & [deg] & [m/s] & [au] &  & [deg] & [deg] & [deg]    \\
        \hline
Mars & All & 305.1$\pm$1.4 & 142$\pm$24 & -42$\pm$11 & 767$\pm$129 & 0.4$\pm$5.2 & 262.7$\pm$4.5 & 312.5$\pm$1.1 & -13.3$\pm$0.6 & 10.2$\pm$0.2 & 1.682$\pm$0.052 & 0.400$\pm$0.016 & 3.19$\pm$0.01 & 304.0$\pm$3.6 & 90.2$\pm$3.1 \\
Mars & $<$ 500 m/s & 304.0$\pm$0.3 & 137$\pm$9 & -49$\pm$4 & 481$\pm$15 & 12.6$\pm$0.8 & 261.3$\pm$3.2 & 311.9$\pm$0.3 & -14.1$\pm$0.1 & 9.8$\pm$0.1 & 1.661$\pm$0.011 & 0.392$\pm$0.004 & 3.21$\pm$0.00 & 308.8$\pm$0.6 & 84.8$\pm$0.4 \\
Mars & $<$ 8 years & 305.7$\pm$0.3 & 105$\pm$3 & -16$\pm$5 & 908$\pm$57 & -5.3$\pm$1.9 & 267.5$\pm$1.6 & 310.6$\pm$0.4 & -14.4$\pm$0.3 & 10.6$\pm$0.1 & 1.801$\pm$0.010 & 0.438$\pm$0.002 & 3.18$\pm$0.01 & 301.1$\pm$2.0 & 89.6$\pm$2.3 \\
Mars & upper & 305.3$\pm$0.7 & 141$\pm$23 & -40$\pm$10 & 787$\pm$129 & -0.0$\pm$5.2 & 263.3$\pm$3.7 & 312.6$\pm$1.1 & -13.3$\pm$0.6 & 10.3$\pm$0.2 & 1.688$\pm$0.052 & 0.403$\pm$0.016 & 3.19$\pm$0.01 & 303.8$\pm$3.4 & 90.4$\pm$3.1 \\
Mars & nominal & 305.1$\pm$0.7 & 140$\pm$24 & -42$\pm$11 & 775$\pm$132 & 0.4$\pm$5.4 & 263.4$\pm$4.0 & 312.5$\pm$1.2 & -13.3$\pm$0.6 & 10.2$\pm$0.2 & 1.684$\pm$0.054 & 0.401$\pm$0.017 & 3.19$\pm$0.01 & 303.7$\pm$3.6 & 90.4$\pm$3.1 \\
Mars & lower & 304.9$\pm$2.1 & 143$\pm$24 & -44$\pm$11 & 748$\pm$124 & 0.8$\pm$4.9 & 261.7$\pm$5.2 & 312.5$\pm$1.1 & -13.3$\pm$0.6 & 10.2$\pm$0.2 & 1.675$\pm$0.050 & 0.398$\pm$0.016 & 3.19$\pm$0.01 & 304.4$\pm$3.7 & 89.9$\pm$3.2 \\
Mars & 10 cm & 304.9$\pm$0.7 & 142$\pm$24 & -42$\pm$11 & 759$\pm$129 & 0.7$\pm$5.1 & 262.5$\pm$4.7 & 312.4$\pm$1.1 & -13.3$\pm$0.6 & 10.1$\pm$0.2 & 1.679$\pm$0.051 & 0.399$\pm$0.016 & 3.19$\pm$0.01 & 304.0$\pm$3.6 & 90.1$\pm$3.2 \\
Mars & 0.5 cm & 305.0$\pm$0.7 & 142$\pm$24 & -42$\pm$11 & 763$\pm$130 & 0.6$\pm$5.2 & 262.6$\pm$4.6 & 312.4$\pm$1.1 & -13.3$\pm$0.6 & 10.2$\pm$0.2 & 1.679$\pm$0.051 & 0.399$\pm$0.016 & 3.19$\pm$0.01 & 303.9$\pm$3.6 & 90.1$\pm$3.2 \\
Mars & 30 $\mu$m & 305.4$\pm$2.4 & 142$\pm$24 & -42$\pm$11 & 784$\pm$127 & -0.1$\pm$5.1 & 262.9$\pm$4.1 & 312.8$\pm$1.2 & -13.3$\pm$0.6 & 10.4$\pm$0.2 & 1.688$\pm$0.055 & 0.404$\pm$0.017 & 3.19$\pm$0.01 & 304.0$\pm$3.5 & 90.5$\pm$3.0 \\
E-M$^*$ & All & 49.1$\pm$5.4 & 203$\pm$3 & 26$\pm$12 & 1835$\pm$111 & -20.1$\pm$7.0 & 247.9$\pm$9.7 & 307.4$\pm$3.4 & -44.7$\pm$2.1 & 4.7$\pm$0.3 & 1.678$\pm$0.031 & 0.411$\pm$0.008 & 4.28$\pm$0.36 & 347.2$\pm$4.3 & 53.6$\pm$5.7 \\
E-M$^*$ & $<$ 20 years & 48.2$\pm$4.7 & 203$\pm$2 & 27$\pm$11 & 1835$\pm$112 & -20.6$\pm$7.0 & 248.5$\pm$9.3 & 307.2$\pm$3.4 & -44.9$\pm$1.8 & 4.6$\pm$0.3 & 1.682$\pm$0.028 & 0.412$\pm$0.008 & 4.32$\pm$0.32 & 347.4$\pm$3.6 & 53.0$\pm$4.8 \\
E-M$^*$ & $<$ 8 years & 41.8$\pm$2.0 & 203$\pm$3 & 49$\pm$4 & 1917$\pm$55 & -31.3$\pm$7.4 & 251.1$\pm$5.3 & 307.0$\pm$2.3 & -48.3$\pm$0.8 & 4.5$\pm$0.2 & 1.748$\pm$0.012 & 0.429$\pm$0.004 & 4.92$\pm$0.14 & 351.7$\pm$1.0 & 45.3$\pm$1.4 \\
        \hline
    \end{tabular}
\end{sidewaystable}

In \citet{Wiegert2020PSJ}, ejecta were modeled as massless particles released from a spherical shell located 1 km from Didymos, with relative velocity directions chosen randomly on a sphere. Three specific ejecta velocities were examined: 10, 100, and 1000 m/s, and the integrations were extended for 10,000 years. The delivery time for meteoroids to near-Earth space was estimated based on the time it takes for the Minimum Orbit Intersection Distance (MOID) to drop to a suitable value. Very small particles (10 $\mu$m) were found to come much closer to Earth in the long term, though our study has not yet sampled particles of that size or duration. Their results indicate that the smallest particles arrive after thousands of years, while larger particles may not arrive at all, except in cases of the highest ejection velocity ($>$1 km/s) where the MOIDs almost immediately reach low values. Higher ejection velocities show that all sizes of particles begin to arrive 15–30 days after the DART impact, but only at ejecta velocities of 6 km/s or more. \citet{Wiegert2020PSJ} found that for ejection velocity above 1 km/s, all sizes except 10 $\mu$m arrive 2 years later. In test simulations where we varied the cone direction, we also achieved a 2-year delivery for the fast-random simulation. Overall, our results are in good agreement with this work.

\begin{figure*}
\centering
\includegraphics[width=0.9\linewidth]{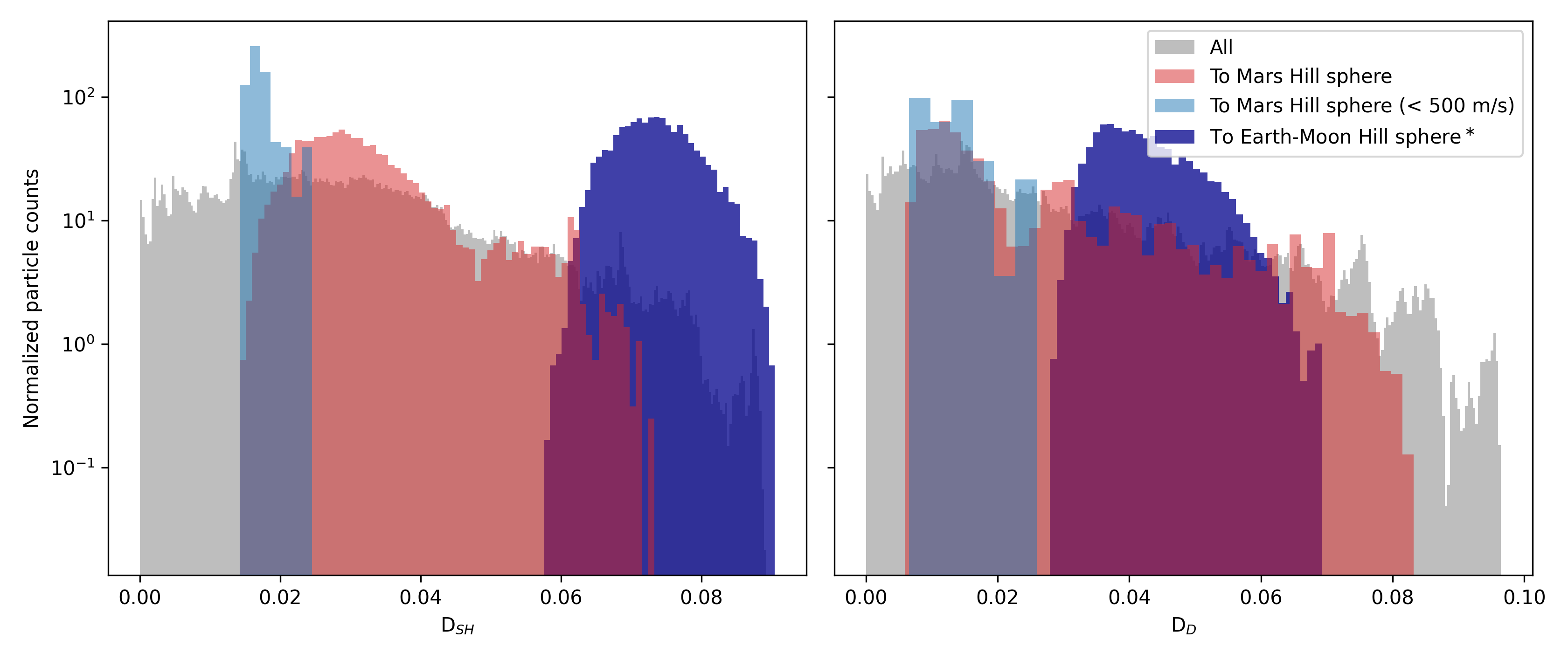}
\caption{Histograms depicting the results of two orbital dissimilarity criteria applied to the last orbital elements of particles in the main simulation. The fast-rnd simulation is denoted by $^*$.
\label{fig:dissimilarity}}
\end{figure*}

To provide practical guidance for identifying DART-created meteors, we analyzed two commonly used orbital dissimilarity criteria mentioned in Section \ref{sec:method}. Figure \ref{fig:dissimilarity} presents histograms that illustrate the outcomes of these criteria when applied to the final orbital elements of particles in the main simulation, and those for impacting the Earth-Moon system for high-speed launch, all concerning the Dimorphos orbital elements at each time. When comparing the two similarity criteria, we observe similar results with the exception of the delivery to the Hill sphere of the Earth-Moon system. The $D_D$ criterion does not exhibit as much decoherence in this case as $D_{SH}$ does. One would expect a larger divergence in the orbital elements because the fast-rnd simulation launches the ejecta at twice the speed. Additionally, the typical cut-off value for $D_D$ is usually 0.05. This implies that 17\% of the DART-created meteors on Mars and 11\% of those on Earth could not be correctly associated with Dimorphos using this criterion. In contrast, $D_{SH}$, which typically has a threshold of 0.2, would allow the correct association of all meteors.

For the delivery of ejecta to Earth within the first 20 years, we also computed the orbit similarity of the potentially DART-created meteors with other known objects and meteoroid streams. The $\alpha$-Aquariids meteor shower shows the highest similarity, with $D_{SH}=0.2$. Regarding near-Earth objects, asteroids 2005 WK56, 2009 UZ19, 2014 TP57, 2020 XU4, 2009 WX7, 2017 VC, 2012 UF, 2018 TQ3, and 2016 TD11 could exhibit lower $D_{SH}$ values than Didymos at the time of impact. Additionally, we calculated the similarity with all meteors observed by the Global Meteor Network (GMN), which has recorded over a million orbits \citep{Vida2021MNRAS5065046V}. We found that only 82 events have $D_{SH}<0.1$. Therefore, should DART-created meteors occur on Earth in the next few years, they would be easily identifiable.


\section{Conclusions} \label{sec:conclusions}

In this study, we utilized a robust simulation framework to model the ejecta produced by the DART impact and investigate the behavior of millions of particles that escaped the Didymos system over a 100-year period. The simulated particles were categorized into three size populations (10 cm, 0.5 cm, and 30 $\mu$m), and solar radiation effects were considered. We constrained the elliptical ejecta cone using early post-impact LICIACube observations. Our results indicate that delivery to Mars could occur at launch velocities above 400 m/s, with the fastest particles, launched at 770 m/s, reaching Mars 7 years after impact. Delivering material to the Earth-Moon system would necessitate higher launch velocities. We determined that the earliest delivery to the Earth-Moon system would be expected at approximately 7 years for ejecta with initial velocities above 1.5 km/s, targeting the Earth-Moon Hill spheres (in no case posing a threat to our planet).

The direction of the ejecta cone slightly influences the delivery process, as well as the aperture angles of the cone, although always in the same order of magnitude. We observed distinct biases in the ejecta directions, directly impacting the launching locations on the surface of Dimorphos. Specifically, we found a predominance of northern launch locations for material delivery to Mars, while delivery to the Earth-Moon system exhibited a bias towards southwestern launch locations. Particle size also played a role in material delivery, with larger particles having a higher probability of reaching Mars, while smaller particles demonstrated a greater propensity for delivery to the Earth-Moon system. Although the effect of particle size was not highly pronounced, it underscores the importance of considering particle size when assessing potential material delivery to different celestial bodies.

The initial location of the ejecta did not affect the fate of particles in our simulation. However, in more complex models incorporating realistic size-velocity-position-direction distributions, this factor will become significant. The temporal features of the delivery, characterized by regular jumps coinciding with orbital periods, bounded solar longitudes of the delivery, and the patterns of the semi-major axis and eccentricity, indicate that the delivery is driven by synodic periods.

To support future observational campaigns, we provided comprehensive information on the encounter characteristics of the delivered material, including well-defined narrow clusters of relative velocity, radiant positions, and orbital elements for both Mars and the Earth-Moon system. This detailed data will aid in the identification of DART-created meteors, enabling researchers to accurately analyze and interpret impact-related phenomena.

We assessed the expected orbital dissimilarity values using multiple criteria to facilitate the identification of DART-created meteors. These criteria offer practical guidance for distinguishing these meteors from other celestial objects, enhancing the accuracy of meteor classification, and enabling targeted studies of DART-related impact events. Based on the widely used threshold values, our analysis indicates that the $D_{SH}$ would be more suitable than $D_D$.

Overall, the DART and Hera missions provide a unique opportunity to study the delivery of impact ejecta to other celestial bodies, offering valuable insights into impact dynamics and the potential recording of DART-created meteors.

\section*{Acknowledgements}
      EP-A acknowledges support from ESA through the Faculty of the European Space Astronomy Centre (ESAC) - Funding reference 149. EP-A also thanks financial support by the LUMIO project funded by the Agenzia Spaziale Italiana (2024-6-HH.0). This project has received funding from the European Research Council (ERC) under the European Union’s Horizon 2020 research and innovation programme (grant agreement No. 865657) for the project “Quantum Chemistry on Interstellar Grains” (QUANTUMGRAIN). JMT-R and EP-A. acknowledge financial support from the project PID2021-128062NB-I00 funded by MCIN/AEI/10.13039/501100011033. AR acknowledges financial support from the FEDER/Ministerio de Ciencia e Innovación – Agencia Estatal de Investigación (PID2021-126427NB-I00, PI: AR). We thank the support received by Prasanna Deshapriya to set the ejecta cone properly.

\bibliography{sample631}{}
\bibliographystyle{aasjournal}

\appendix

\begin{sidewaysfigure}[htbp]
\includegraphics[width=\linewidth]{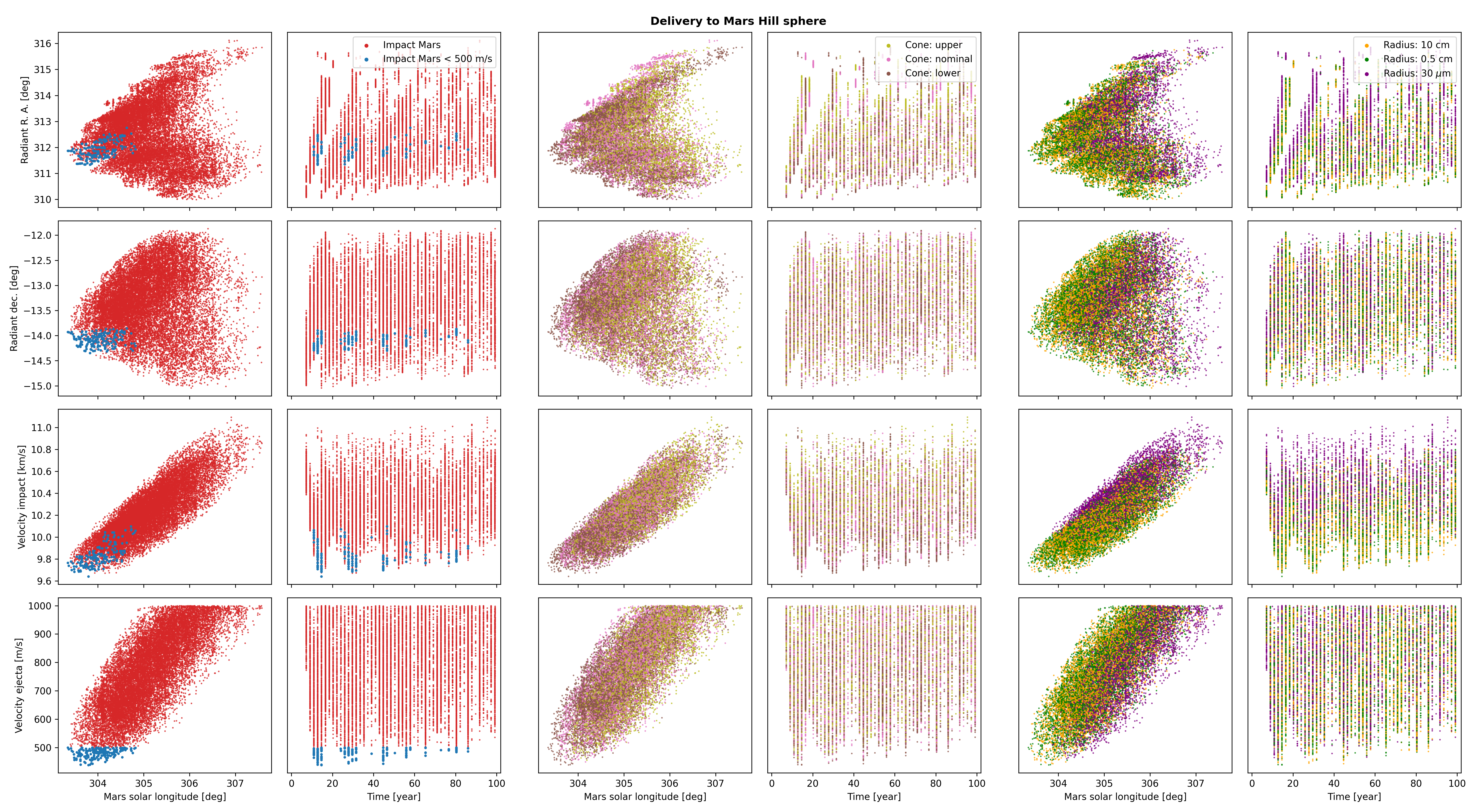}
\caption{Right ascension and declination in the J2000 reference frame centered on Mars, along with the relative impact velocity and the launch velocity in relation to Mars' solar longitude and impact time for the delivered material to Mars of the main simulation. The data is subdivided into (two left columns) particles launched at velocities below 500 m/s, (two middle columns) three initial ejecta cone configurations, and (two right columns) different sizes.
\label{fig:fate_radiant}}
\end{sidewaysfigure}
\begin{sidewaysfigure}[htbp]
\includegraphics[width=\linewidth]{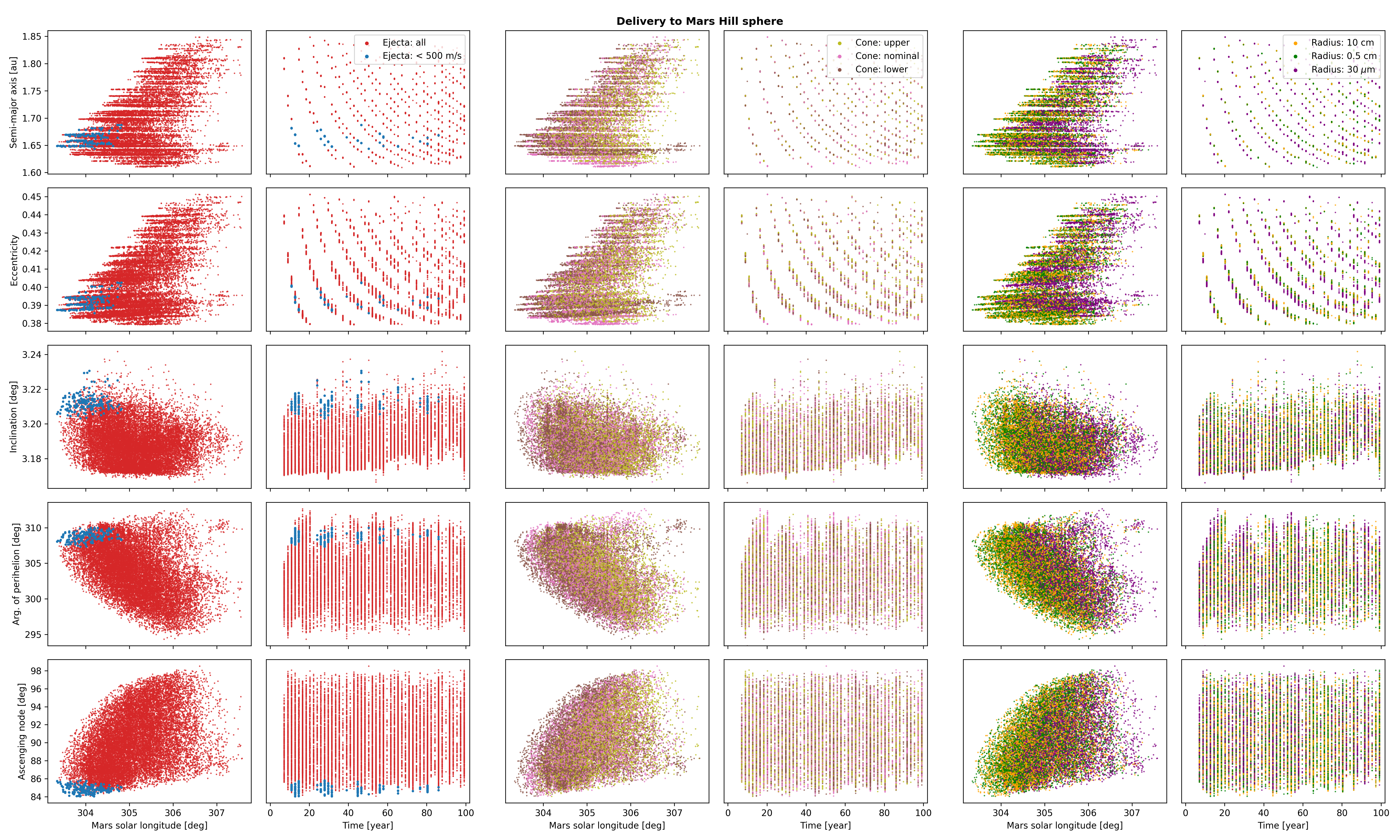}
\caption{Semi-major axis, eccentricity, inclination, argument of the perihelion, and ascending node in the ECLIPJ2000 reference frame in relation to Mars' solar longitude and impact time for the delivered material to Mars of the main simulation. The data is subdivided into (two left columns) particles launched at velocities below 500 m/s, (two middle columns) three initial ejecta cone configurations, and (two right columns) different sizes.
\label{fig:fate_orbelem}}
\end{sidewaysfigure}
\begin{sidewaysfigure}[htbp!]
\includegraphics[width=\linewidth]{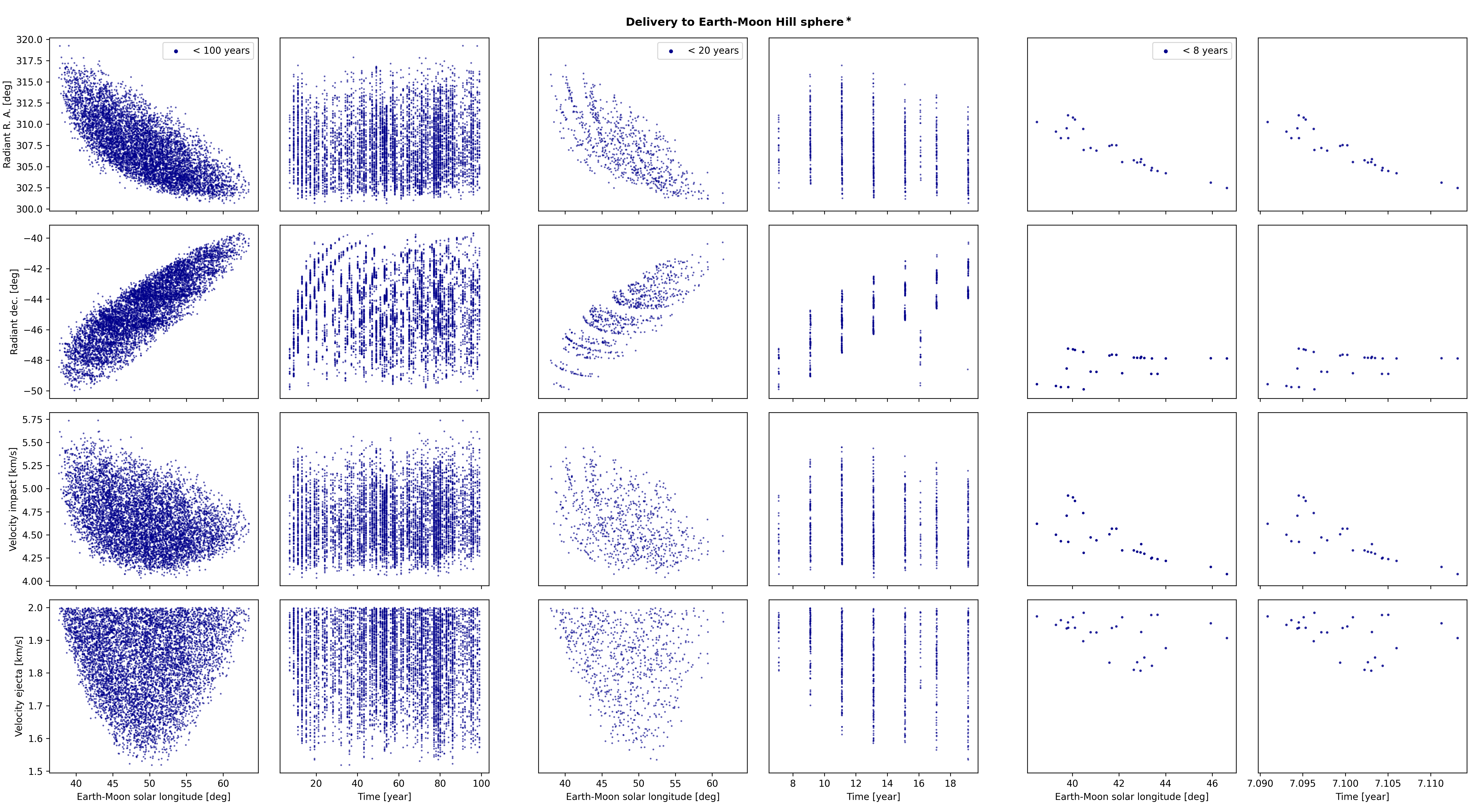}
\caption{Right ascension and declination in the J2000 reference frame centered on Earth-Moon, along with the relative impact velocity and the launch velocity in relation to Earth-Moon's solar longitude and impact time for the delivered material to Mars of the simulation with launch velocity ranging from 1 to 2 km/s, randomly distributed along the impact crater. The data is subdivided into (two left columns) all particles delivered over 100 years, (two middle columns) over 20 years, and (two right columns) over 8 years.
\label{fig:earth_fate_radiant}}
\end{sidewaysfigure}
\begin{sidewaysfigure}[htbp!]
\includegraphics[width=\linewidth]{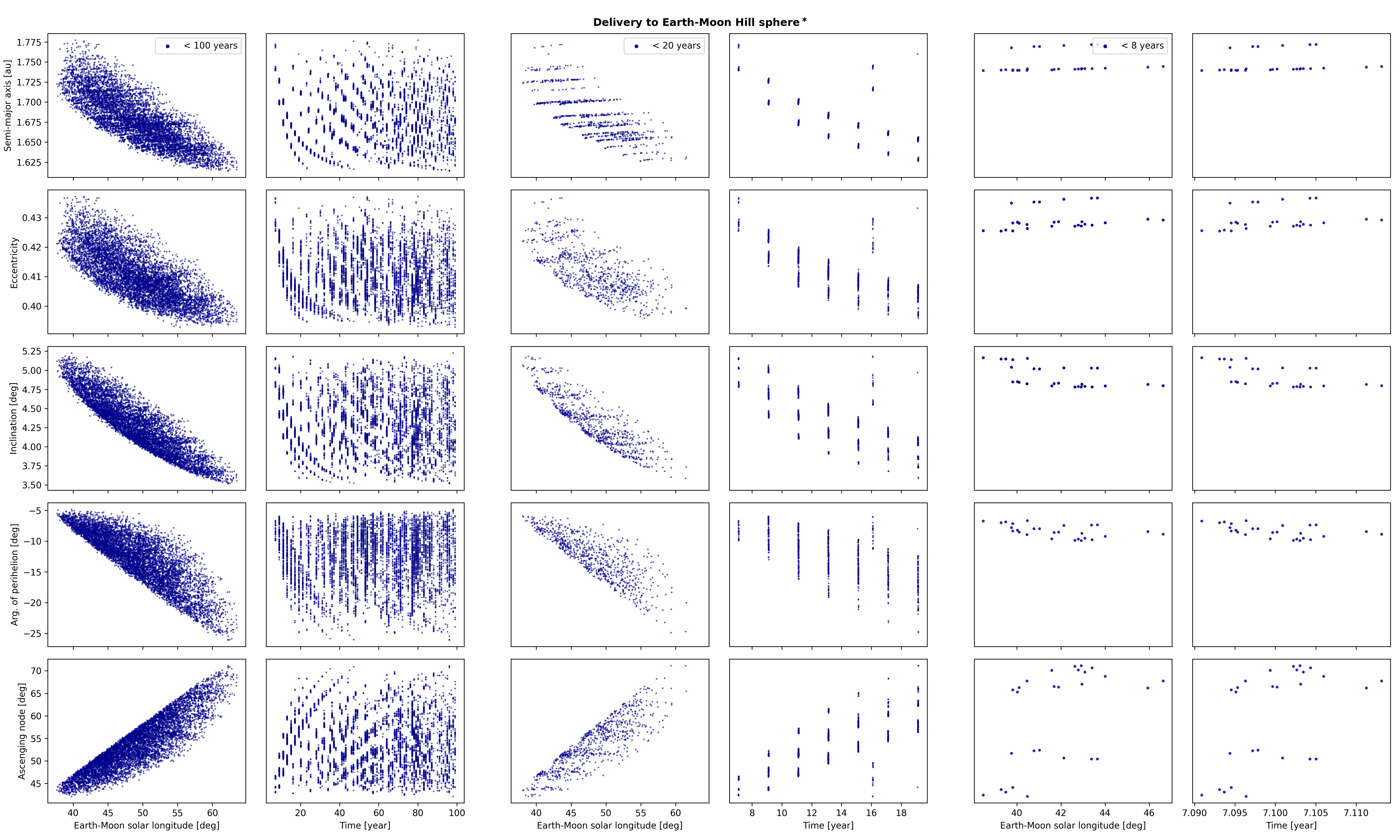}
\caption{Semi-major axis, eccentricity, inclination, argument of the perihelion, and ascending node in the ECLIPJ2000 reference frame in relation to Earth-Moon's solar longitude and impact time for the delivered material to Earth-Moon of the simulation with launch velocity ranging from 1 to 2 km/s, randomly distributed along the impact crater. The data is subdivided into (two left columns) particles delivered over 100 years, (two middle columns) over 20 years, and (two right columns) over 8 years.
\label{fig:earth_fate_orbelem}}
\end{sidewaysfigure}

This appendix includes detailed figures providing the distribution of some parameters of the delivered particles to Mars and Earth Hill spheres. Figure \ref{fig:fate_radiant} shows the right ascension and declination in the J2000 reference frame centered on Mars, along with the relative impact velocity and launch velocity relative to Mars' solar longitude and impact time for material delivered to Mars in the main simulation. The data is categorized into particles launched at velocities below 500 m/s (left columns), four temporal impact windows (middle columns), and different sizes (right columns). Figure \ref{fig:fate_orbelem} illustrates the semi-major axis, eccentricity, inclination, argument of the perihelion, and ascending node in the ECLIPJ2000 reference frame in relation to Mars' solar longitude and impact time for the material delivered to Mars in the main simulation. The data is divided into particles launched at velocities below 500 m/s (left columns), four temporal impact windows (middle columns), and different sizes (right columns). Figure \ref{fig:earth_fate_radiant} presents the right ascension and declination in the J2000 reference frame centered on Mars, along with the relative impact velocity and launch velocity for material delivered to Mars. This simulation considers launch velocities ranging from 1 to 2 km/s, randomly distributed along the impact crater. The data is categorized into particles delivered over 100 years (left columns), 20 years (middle columns), and 8 years (right columns). Figure \ref{fig:earth_fate_orbelem} shows the semi-major axis, eccentricity, inclination, argument of the perihelion, and ascending node in the ECLIPJ2000 reference frame for material delivered to Mars. This simulation uses launch velocities ranging from 1 to 2 km/s, randomly distributed along the impact crater. The data is divided into particles delivered over 100 years (left columns), 20 years (middle columns), and 8 years (right columns).

\end{document}